\documentclass[ste,twoside]{stefano}

\usepackage{pstricks}
\usepackage{amsmath}
\usepackage{amssymb}
\usepackage{graphics}
\usepackage{graphicx}
\usepackage{epsfig}
\usepackage{rotating}
\usepackage{cite}
\usepackage{color}

\def\makeheadbox{{%
\hbox to0pt{\vbox{\baselineskip=10dd\hrule\hbox
to\hsize{\vrule\kern3pt\vbox{\kern3pt \hbox{  {\sc J. Math. Phys.} {\bf
47}, 082106-15 (2006) } \hbox{ {\sc
{\color{blue}{dma}}[{\color{black}{imecc}}]{\color{red}{UniCamp}} }
\hspace*{10.4cm} {\color{blue}{$\boldsymbol{\Sigma \delta \Lambda}$}} }
\kern3pt}\hfil\kern3pt\vrule}\hrule}%
\hss}}}

\def\r{\boldsymbol{r}}

\def\I{\mbox{\tiny I}}
\def\+{\mbox{\tiny $\dag$}}
\def\0{\mbox{\tiny $0$}}
\def\1{\mbox{\tiny $1$}}
\def\2{\mbox{\tiny $2$}}
\def\3{\mbox{\tiny $3$}}
\def\4{\mbox{\tiny $4$}}
\def\5{\mbox{\tiny $5$}}
\def\6{\mbox{\tiny $6$}}
\def\7{\mbox{\tiny $7$}}
\def\8{\mbox{\tiny $8$}}
\def\9{\mbox{\tiny $9$}}

\def\Sh{\mbox{\tiny $S$}}
\def\B{\mbox{\tiny $B$}}
\def\C{\mbox{\tiny $C$}}

\def\t{\mbox{\tiny $t$}}

\def\pem{\mbox{\tiny $\pm$}}
\def\mi{\mbox{\tiny $-$}}
\def\pl{\mbox{\tiny $+$}}
\def\={\mbox{\tiny $=$}}
\def\T{\mbox{\tiny $T$}}
\def\L{\mbox{\tiny $L$}}
\def\R{\mbox{\tiny $R$}}

\def\Co{\mbox{\tiny $\mathbb{C}$}}

\def\Sh{\mbox{\tiny S}}
\def\A{\mbox{\tiny $A$}}
\def\B{\mbox{\tiny $B$}}
\def\C{\mbox{\tiny $C$}}
\def\I{\mbox{\tiny I}}
\def\II{\mbox{\tiny II}}
\begin{document}
%

\title{ANALYTIC PLANE WAVE SOLUTIONS FOR THE QUATERNIONIC POTENTIAL STEP}

\author{
Stefano De Leo\inst{1}
\and Gisele C. Ducati\inst{2} \and Tiago M. Madureira\inst{2}
}

\institute{
Department of Applied Mathematics, University of Campinas\\
PO Box 6065, SP 13083-970, Campinas, Brazil\\
{\em deleo@ime.unicamp.br}
 \and
Department of Mathematics, University of Parana\\
PO Box 19081, PR 81531-970, Curitiba, Brazil\\
{\em ducati@mat.ufpr.br}\\
{\em tmadureira@mat.ufpr.br}
}


\date{Submitted: {\em April, 2006}. Revised: {\em May, 2006}.   }

\abstract{By using the recent mathematical tools developed in
quaternionic differential operator theory, we solve the
Schr\"odinger equation in presence of a quaternionic step
potential. The analytic solution for the stationary states allows
to explicitly show the qualitative and quantitative differences
between this quaternionic quantum dynamical system and its complex
counterpart. A brief discussion on reflected and transmitted
times, performed by using the stationary phase method, and its
implication on the experimental evidence for deviations of
standard quantum mechanics is also presented. The analytic
solution given in this paper represents a fundamental mathematical
tool to find an analytic approximation to the quaternionic barrier
problem (up to now solved by numerical method).}












\titlerunning{\sc quaternionic plane wave solutions for the potential step}

\maketitle


\section*{I. INTRODUCTION}

Since a quaternionic equation can be equivalently written as a
two-component complex equation, it is natural to ask whether the
quaternionic Schr\"odinger equation is simply another way of
rewriting complex quantum mechanics. The answer to this question
is given in his famous book\cite{ADL} by Adler. Probabilities in
quaternionic dynamical system are different from those of standard
complex theory. In the first papers on the quaternionic
Schr\"odinger equation deviations from complex quantum mechanics
were studied by considering quaternionic perturbation
potentials\cite{DAV89,DAV92}. Recent progresses on the solution of
quaternionic differential equations\cite{DEDUC1,DEDUC1a,DEDUC1b}
have improved the physical discussion on quaternionic tunnelling
phenomena\cite{DEDUC2} and bound states\cite{DEDUC3}.

In this paper, an interesting simple quaternionic quantum
mechanical system is {\em analytically} solved. This allows to
discuss both qualitative and quantitative differences between
quaternionic and complex quantum mechanics. The explicit
stationary wave solution for the quaternionic potential step shows
some important results which could be very useful in looking for
deviations from the standard quantum theory. For example, the
quaternionic step diffusion is characterized by reflected and
transmitted waves which are {\em not} instantaneous. The analytic
solution is also very useful to understand the effect that
quaternionic potentials play on the phase of stationary waves. The
advantage to analytically solve a quaternionic problem is surely
represented by the possibility to  deeply study the quaternionic
solution and understand where and if deviations from complex
quantum theory could be seen.

\section*{II. QUATERNIONIC SCHR\"ODINGER EQUATION}

In the quaternionic formulation of non-relativistic quantum
mechanics, the dynamics of a particle without spin subject to the
influence of the anti-hermitian scalar potential,
\[ i\,V_{\1}(\r, t) + j\,V_{\2}(\r, t) +k\,V_{\3}(\r, t)\,\,, \]
is described by
\begin{equation}
\label{qse} \hbar\,\partial_{\t} \Psi( \r,t) = \left[ \,
i\,\mbox{$\frac{\hbar^{^{\2}}}{2m}$} \, \nabla^{\2} - i\,
V_{\1}(\r, t) - j \, V_{\2}(\r, t) - k \, V_{\3}(\r, t) \, \right]
\, \Psi(\r,t)~,
\end{equation}
with
\[
V_{\1,\2,\3}: (\mathbb{R}^{\3},\mathbb{R}) \to \mathbb{R}
\hspace*{.5cm} \mbox{and} \hspace*{.5cm}  \Psi:
(\mathbb{R}^{\3},\mathbb{R}) \to \mathbb{H}\,\,.\] Eq.(\ref{qse})
is known as the Schr\"odinger equation for quaternionic quantum
mechanics\cite{ADL}. It is natural to try to relate the new
results coming from this quaternionic formulation with the well
know phenomena discussed in the standard textbooks of (complex)
quantum mechanics\cite{COHEN,MAT,MER}. In this spirit, the complex
limit, i.e. $V_{\2,\3} \to 0$,  surely represents a useful
mathematical tool to test quaternionic calculations and to
understand, by explicitly showing the difference between the
quaternionic and complex formulation, if and where quaternionic
deviations from standard quantum mechanics could be seen and
investigated.

The linearity in $\partial_t$ of the evolution time operator in
Eq.(\ref{qse}) guarantees to obtain a positive probability density
\begin{equation}
\rho(\r, t)=\overline{\Psi}(\r, t) \Psi(\r, t)\,\,
,\end{equation}
 together with a continuity equation
\begin{equation}
\label{con}
\partial_{\t} \rho(\r, t) +
\boldsymbol{\nabla} \cdot \boldsymbol{J}(\r, t) = 0~.
\end{equation}
To find the explicit form of the current density
$\boldsymbol{J}(\r, t)$, let us first derive the Schr\"odinger
equation for $\overline{\Psi}(\r, t)$ [the quaternionic conjugate
of $\Psi(\r, t)$, i.e. $(i,j,k)\to - (i,j,k)$],
\begin{equation}
\label{qsec}
\hbar\,
\partial_{\t} \overline{\Psi}( \r,t) =
 - \mbox{$\frac{\hbar^{\2}}{2m}$} \, \nabla^{^{\2}}
\overline{\Psi}( \r,t)\, i + \overline{\Psi}( \r,t) \left[\, i\,
V_{\1}(\r, t) + j \, V_{\2}(\r, t) + k \, V_{\3}(\r, t) \,
\right]~.
\end{equation}
Combining Eq.(\ref{qse}) [multiplied from the left by
$\overline{\Psi}( \r,t)$] and Eq.(\ref{qsec}) [multiplied from the
right by  $\Psi( \r,t)$], we obtain
\[
\partial_t\left[\overline{\Psi}( \r,t)\Psi( \r,t)\right] + \,
\mbox{$\frac{\hbar}{2m}$}\, \left\{\left[ \nabla^{^{\2}}
\overline{\Psi}( \r,t)\right]  \,i\, \Psi( \r,t) -
\overline{\Psi}( \r,t) \,i\, \nabla^{^{\2}} \Psi( \r,t)\right\} =
0\,\,.
\]
Consequently, the density current in quaternionic quantum
mechanics is formally equal to that one of the usual complex
theory, i.e.
\begin{equation}
\boldsymbol{J}( \r,t) = \mbox{$\frac{\hbar}{2m}$} \, \left\{\left[
\nabla  \overline{\Psi}( \r,t)\right]  \,i\, \Psi( \r,t) -
\overline{\Psi}( \r,t) \,i\, \nabla  \Psi( \r,t)\right\}\,\,.
\end{equation}
It is worth pointing out that, due to the non commutativity nature
of quaternions, the position of the imaginary unit $i$  is not a
choice but it is imposed by the anti-hermiticity of evolution time
operator in Eq.(\ref{qse}).

\subsection*{$\bullet$ TIME INDEPENDENT POTENTIALS}

In this paper, we are going to be concerned with a particle in a
time independent potential. In complex quantum mechanics, the
rapid spatial variations of a square potential introduce purely
quantum effects in the motion of the particle. The same is valid
for perturbative quaternionic potentials. Before beginning our
investigation, we shall discuss some important mathematical
properties of the quaternionic Schr\"odinger equation in the
presence of time independent potentials,
\begin{equation} \label{qset}
\hbar\,\partial_{\t} \Psi( \r,t) = \left[ \,
i\,\mbox{$\frac{\hbar^{^{\2}}}{2m}$} \, \nabla^{\2} - i\,
V_{\1}(\r) - j \, V_{\2}(\r) - k \, V_{\3}(\r) \, \right] \,
\Psi(\r,t)~.
\end{equation}
Taking into account that $\Psi(\r, t)$ is a quaternionic function,
we apply the method of separation of variables with the time
dependent function appearing in the right hand side\cite{DEDUC1},
\begin{equation}
\label{ss} \Psi(\r, t) = \Phi(\r) \, \exp[ \, - \,
\mbox{$\frac{i}{\hbar}$} \, E \, t \, ]\,\,,
\end{equation}
with
\[\Phi:\mathbb{R}^{\3} \to \mathbb{H}\,\,.\]
 This stationary solution of the Schr\"odinger equation leads
to a time-independent probability density $\rho(\r)$.
Consequently, the current density satisfies
\begin{equation}
\label{cd2} \boldsymbol{\nabla} \cdot \left\{\,\left[ \nabla
\overline{\Phi}( \r)\right]  \,i\, \Phi( \r) - \overline{\Phi}(
\r) \,i\, \nabla  \Phi( \r)\,\right\}  = 0~.
\end{equation}
By using the separation of variable (\ref{ss}), Eq.(\ref{qset})
reduces to the following quaternionic (right) eigenvalue
problem\cite{DESCO1,DESCO2},
\begin{equation} \label{qse2} \left[ \, i \,
\mbox{$\frac{\hbar^{\2}}{2m}$} \, \nabla^{\2} - i \, V_{\1}(\r) -
j \, V_{\2}(\r) - k \, \, V_{\3}(\r) \right] \, \Phi(\r) +
\Phi(\r) \, i \, E = 0~.
\end{equation}

\subsection*{$\bullet$  TIME REVERSAL INVARIANCE}

From Eq.~(\ref{qset}), we can immediately obtain  the
time-reversed Schr\"odinger equation
\begin{equation}
\label{tr} \hbar \,\partial_{\t} \Psi_{\T}( \r, - t) = - \left[ \,
i\,\mbox{$\frac{\hbar^{^{\2}}}{2m}$} \, \nabla^{\2} - i\,
V_{\1}(\r) - j \, V_{\2}(\r) - k \, V_{\3}(\r) \, \right] \,
\Psi_{\T}(\r,-t)~.
\end{equation}
In complex quantum mechanics the $*$-conjugation yields a
time-reversed version  of the original Schr\"odinger equation. In
quaternionic quantum mechanics there does {\em not} exist a
universal time reversal operator~\cite{ADL}. Only a {\em
restricted} class of time-independent quaternionic potentials
$[V_{\2}(\r)\propto V_{\3}(\r)]$, i.e.
\[
W(\r)=V_{\2}(\r) - i\,V_{\3}(\r) = |W(\r)| \, \exp [\,i\,
\theta\,]\,\,\,\,\,\,\,\,\,(\theta=\mbox{const})~,
\]
is time reversal invariant.  For these potentials,
\begin{equation}
\Psi_{\T}(\r, - t) =  u \,   \Psi(\r, t) \, \bar{u}~,~~~ u = k \,
\exp \left[ \,i \, \theta \,\right]~.
\end{equation}
In the standard quantum mechanics limit, due to the complex nature
of the wave function $\Psi_{\Co}(\r, t)$, we recover the well
known result
\[\Psi_{\Co,\T}(\r, - t) = \Psi_{\Co}^{*}(\r, t)\,\,.\]

\subsection*{$\bullet$ ONE-DIMENSIONAL SQUARE POTENTIALS}

Let us consider one-dimensional potentials. In the case of square
shapes, the potential is a quaternionic constant in certain
regions of space. In such regions, the stationary wave function
$\Phi(x)$ is obtained by solving the following second order
differential equation with (left) constant quaternionic
coefficients,
\begin{equation}
\label{qse3}
 \left[ i \, \mbox{$\frac{\hbar^{\2}}{2m}$} \, \Phi''(x)
 - i \, V_{\1} - j \, V_{\2} - k \, V_{\3} \, \right] \, \Phi(x) =
 -\,
\Phi(x) \, i \, E ~.
\end{equation}
It is not our purpose here to discuss the theory of quaternionic
differential equations and we refer the interested reader to the
papers cited in refs.\cite{DEDUC1,DEDUC1a,DEDUC1b} where a
detailed exposition of the subject is found. The solution of
Eq.(\ref{qse3}) is
\begin{equation}
\label{solg} \Phi(x) =   ( 1 + j w) \,  \left\{ \exp \left[ \,
\nu_{\mi} \, x \, \right] \, c_{\1} + \exp \left[ \, - \,
\nu_{\mi} \, x \, \right] \, c_{\2} \right\} +  ( z + j ) \,
\left\{ \exp \left[ \, \nu_{\pl}  \, x \, \right] \, c_{\3} + \exp
\left[ \, - \, \nu_{\pl} \, x \, \right] \, c_{\4} \right\}~,
\end{equation}
where $c_{\1,...,\4}$ are complex coefficients to be determined by
the boundary conditions and
\[
\begin{array}{ccl}
\nu_{\pem} & = & \sqrt{ 2m\, \left( V_{\1} \pm \sqrt{E^{\2} -
V_{\2}^{\2} - V_{\3}^{\2}} \right)} \,/\,\hbar \,\,,\\ \\ z & =&
i \,\left(V_{\2} + i V_{\3}\right) \,/\, \left( E + \sqrt{E^{\2} -
V_{\2}^{\2} - V_{\3}^{\2}} \right) \,\,,\\ \\w & =& - \,i
\,\left(V_{\2} - i V_{\3}\right) \,/\, \left( E + \sqrt{E^{\2} -
V_{\2}^{\2} - V_{\3}^{\2}} \right) \hspace*{.5cm}\in
\mathbb{C}(1,i)~.
\end{array}
\]
In the free potential region ($V_{\1,\2,\3} = 0$) the previous
solution reduces to
\begin{eqnarray}
\label{free}
\Phi(x) & = & \exp \left[ \, i \, \epsilon \, x \,
\right] \, c_{\1} + \exp \left[ \, - \, i \,  \epsilon  \, x \,
\right] \, c_{\2} + j \, \left\{ \exp \left[ \,
 \epsilon  \, x \, \right] \, c_{\3} +
 \exp \left[ \, - \,
 \epsilon  \, x \,  \right] \, c_{\4}
\right\}~,
\end{eqnarray}
where \[\epsilon = \sqrt{2mE}/\hbar\,\,\,\in \mathbb{R}\,\,.\]


\section*{III. BOUNDED SOLUTIONS AND CURRENT DENSITY}

Let us now calculate the stationary states in the case of a
quaternionic step potential. The procedure follows the standard
one. We use Eq.(\ref{solg})  in the region where the potential is
a constant and Eq.(\ref{free}) in the free region. We then impose
that such solutions remain bounded and, finally, we match these
functions by requiring the continuity of $\Phi(x)$ and its
derivative in $x=0$. Before proceeding  with our calculations, the
only point deserving further discussion concerns the
classification of the energy zones in the potential region (in
order to distinguish between partial and total reflection). To do
this, we have to analyze the complex exponential factors\,
$\nu_{\pem}$. The possible cases are sketched in the following
figure
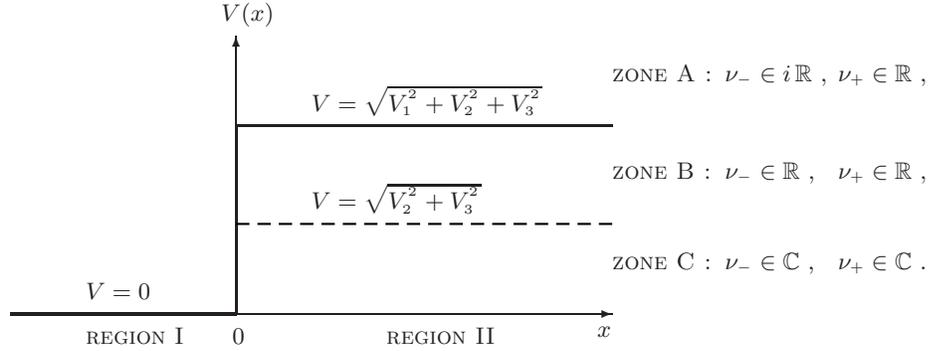
\begin{center}
\setlength{\unitlength}{1cm}
\begin{picture}(10,4.5)
\put(1,0.5){\small$V=0$} \put(1,-0.1){\small\textsc{region I}}
\put(5,-0.1){\small\textsc{region II}} \put(2.8,4.2){\small$V(x)$}
\put(7.8,0){\small$x$} \put(4,3.0){\small$V=\sqrt{V_{\1}^{^{\2}} +
V_{\2}^{^{\2}} + V_{\3}^{^{\2}}}$} \put(4,1.7){\small
$V=\sqrt{V_{\2}^{^{\2}} + V_{\3}^{^{\2}}}$}
\put(2.95,-0.1){\small$0$} \put(0,0.3){\vector(1,0){8}}
\put(3,0.3){\vector(0,1){3.7}} \thicklines
\put(0,0.3){\line(1,0){3}} \put(3,0.3){\line(0,1){2.5}}
\psline[linestyle=dashed](3,1.5)(8,1.5)
\put(3,2.8){\line(1,0){5}}\put(8,3.4){\small\textsc{zone A :}}
\put(8,2.1){\small\textsc{zone B :}}\put(8,0.9){\small
\textsc{zone C :}} \put(9.5,3.4){\small $\nu_{\mi}\in i\,
\mathbb{R}\,\,,$} \put(11.,3.4){\small $\nu_{\pl}\in
\mathbb{R}\,\,,$} \put(9.5,2.1){\small $\nu_{\mi}\in
\mathbb{R}\,\,,$} \put(11.,2.1){\small $\nu_{\pl}\in
\mathbb{R}\,\,,$} \put(9.5,.9){\small $\nu_{\mi}\in
\mathbb{C}\,\,,$} \put(11.,.9){\small $\nu_{\pl}\in
\mathbb{C}\,\,.$}
\end{picture}
\end{center}

\vspace*{.8cm}

\noindent To avoid any confusion between real and imaginary
coefficients and to facilitate the reading of this paper, in the
sequel, we shall adopt  the following notation
\[
\begin{array}{lll}
\mbox{\sc zone A :}\,\,\,\,\, & \nu_{\mi} = i \,\rho_{\mi}
\,\,,\,\,\,\,\,
& \nu_{\pl}\,\,, \\
\mbox{\sc zone B :}\,\,\,\,\, & \nu_{\mi} \,\,,\,\,\,\,\,
& \nu_{\pl}\,\,, \\
\mbox{\sc zone C :}\,\,\,\,\, & \nu_{\mi}=\sigma_{\pl} -
i\,\sigma_{\mi}  \,\,,\,\,\,\,\, & \nu_{\pl}=\sigma_{\pl} +
i\,\sigma_{\mi}\,\,,
\end{array}
\]
where
\[
\rho_{\mi} =
\sqrt{\dfrac{2\,m}{\hbar^{^{\2}}}\,\left(\sqrt{E^2-V_2^2-V_3^2} -
V_1\right)}\,\,,\,\,\,
\sigma_\pm=\sqrt{\dfrac{m}{\hbar^{^{\2}}}\,\left(\sqrt{V_1^2+V_2^2+V_3^2-E^2}\pm
V_1\right)}\,\,\,\,\,\in \mathbb{R}\,\,.
\]

\subsection*{$\bullet$ REGION I}

For the solution to remain bounded when $x\to -\infty$, it is
necessary to have $c_{\4}=0$ in Eq.(\ref{free}). So, the solution
in region I becomes
\begin{equation}
\label{frees} \Phi_{\I}(x) =  e^{i\epsilon x} + r \, e^{\mi\,i
\epsilon x} + j\, \tilde{r}\, e^{\epsilon x}\,\,,
\end{equation}
where  $c_{\2}=r$  and $c_{\3}=\tilde{r}$ represent the reflection
coefficients to be determined by the matching conditions. From
Eq.(\ref{cd2}), we immediately find the constant value of the
current density in this region, i.e.
\begin{equation}
\label{j1} J_{\I}=(1-|r|^2)\, \hbar \,\epsilon/m\,\,.
\end{equation}

\subsection*{$\bullet$ REGION II - ZONE A: PARTIAL REFLECTION}

The condition on the boundedness of the solution implies that
$c_{\3} = 0$ in Eq.(\ref{solg}). Since, the incident particle is
coming from $x=-\infty$, we also have to impose $c_{\2} = 0$. The
stationary wave function, in zone A,  is then given by
\begin{equation}
\label{IIA} \Phi_{\II,\A}(x) = ( 1 + jw )\, t\, e^{i \,\rho_{\mi}
x} + ( z + j ) \,\tilde{t} \, e^{- \rho_{\pl} x}~,
\end{equation}
where  $c_{\1}=t$  and $c_{\4}=\tilde{t}$ represent the
transmission coefficients to be determined by the matching
conditions. In this region the current density is
\begin{equation}
J_{\II,\A} = (1-|w|^2) \, |t|^2\, \hbar\, \rho_{\mi}/m\,\,.
\end{equation}
This means a non-null transmission probability and consequently
partial reflection in region I.

\subsection*{$\bullet$ REGION II - ZONE B:  TOTAL REFLECTION}

For the solution to remain bounded when $x\to +\infty$, it is
necessary that $c_{\2}=c_{\4}=0$ in Eq.(\ref{solg}). Thus, the
solution in zone B is
\begin{equation}
\label{IIB}
 \Phi_{\II,\B}(x) =
(1+jw)\,t\,e^{-\nu_-x}+(z+j)\,\tilde{t}\,e^{-\nu_{\pl}x}~.
\end{equation}
In this zone, the current density is null
\begin{equation}
J_{\II,\B}=0~.
\end{equation}
This characterizes a total reflection in region I.

\subsection*{$\bullet$ REGION II - ZONE C: TOTAL REFLECTION}

The boundedness condition of the solution implies that $c_{\1} =
c_{\3} = 0$ in Eq.(\ref{solg}). Thus, we have
\begin{equation}
\label{IIC} \Phi_{\II,\C}(x) = \left[ (1+jw)\, t\, e^{i\,
\sigma_{\mi} x} + ( z + j ) \,\tilde{t} \,e^{-i \,\sigma_{\mi} x}
\right] \, e^{- \sigma_{\pl} x}~,
\end{equation}
with
\[
w = -i \dfrac{V_{\2} - i V_{\3}}{\sqrt{V_{\2}^{^{\2}} +
V_{\3}^{^{\2}}}} e^{-i\varphi}~, \quad z = i \dfrac{V_{\2} + i
V_{\3}}{\sqrt{V_{\2}^{^{\2}}+V_{\3}^{^{\2}}}}e^{-i\varphi}\quad
\mbox{and} \quad \varphi = \arctan \left[
\dfrac{\sqrt{V_{\2}^{^{\2}} + V_{\3}^{^{\2}} - E^{^{\2}}}}{E}
\right]~.
\]
As in the previous zone, the current density is null
\begin{equation}
J_{\II,\C}=0~.
\end{equation}
This implies total reflection in region I.

\subsection*{$\bullet$ RELATION BETWEEN REFLECTION AND
TRANSMISSION COEFFICIENTS} 

The stationary wave solution of the Schr\"odinger equation in the
presence of a quaternionic step potential  can be then expressed
in terms of  complex reflection ($r,\tilde{r}$) and transmission
($t,\tilde{t}$) coefficients:
\begin{eqnarray}
&&\Phi_{\I} (x) = e^{i \epsilon x}+ r \,e^{- \, i \epsilon x} +
j \,\tilde{r}\, e^{\epsilon x}\,\,,\\[3mm]
&&\Phi_{\II}(x) = \left\{ \begin{array}{lr} (1 + j\, w )\, t\,
e^{i \,\rho_{\mi} x} + ( z + j ) \,\tilde{t} \,e^{- \nu_{\pl} x}
&  \mbox{\sc zone A}\,\,,\\[2mm] 
( 1 + j \,w )\, t \,e^{- \nu_{\mi} x} + ( z + j ) \,\tilde{t}
\,e^{-
\nu_{\pl} x} &  \mbox{\sc zone B}\,\,,\\[2mm] 
\left[ ( 1 + j \,w ) \,t \,e^{i \,\sigma_{\mi} x} + ( z + j )\,
\tilde{t} \,e^{- \, i \,\sigma_{\mi} x} \right] \,e^{-
\sigma_{\pl} x} & ~~~~\mbox{\sc zone C}\,\, .
\end{array} \right.
\end{eqnarray}
As we saw, the current density assumes a constant value. This
value   has been calculated in the free potential region and in
each of the three different zones of region II.
 The continuity of $\Phi(x)$ and its derivative in $x=0$ implies
 the continuity of the current density, i.e. $J_{\I} = J_{\II}$. This
 gives an immediate relation between reflection and transmission
 coefficients,
\begin{equation}
R + T = 1\,\,,
\end{equation}
 with
 \[
\begin{array}{lclll}
 R=|r|^2 & \,\,\,\mbox{and}\,\,\,&
T=\dfrac{\rho_{\mi}}{\epsilon}(1-|w|^2)|t|^2 &
\,\,\,\,\,\,\,\,\,\,\mbox{for} &
\,\,E>\sqrt{V_1^2+V_2^2+V_3^2}\,\, ,\\
R=|r|^2 & \,\,\,\mbox{and}\,\,\,& T=0&
\,\,\,\,\,\,\,\,\,\,\mbox{for} &
\,\,E<\sqrt{V_1^2+V_2^2+V_3^2}\,\, .
\end{array}
\]
Observe that, both in complex and quaternionic quantum mechanics,
to find the relation between $R$ and $T$ we do not have the
necessity to find the explicit value of plane wave coefficients
$r$ and $t$.

 %

\section*{IV. EXPLICIT PLANE WAVE SOLUTIONS}

 The usual method for determining the
stationary states in a square potential requires the continuity of
$\Phi(x)$ and its derivative at the point where the potential is
discontinuous (in this case $x=0$). Then, we impose that

 \begin{eqnarray}
&&\Phi_{\I}(0)=\Phi_{\II}(0)   \nonumber \\
&&\Phi_{\I}'(0)=\Phi_{\II}'(0)~.
\end{eqnarray}

\subsection*{$\bullet$ REGION II - ZONE A: CONTINUITY}

Matching the conditions at $x = 0$, we get
 \begin{eqnarray*}
1 + r + j\, \tilde{r} &=& ( 1 + j\,w)\, t +
( z + j ) \,\tilde{t}~, \\
i \,\epsilon \,( 1 - r ) + j \,\epsilon \,\tilde{r} &=& ( 1 + j\,w
)\, i \,\rho_{\mi}\,t -(z+j)\,\nu_{\pl}\,\tilde{t}~.
\end{eqnarray*}
After separating the complex from the pure quaternionic part, we
find
\[
\begin{array}{l}
1 + r = t + z \,\tilde{t}\,\,,\\[2mm]
\tilde{r} = w \,t + \tilde{t}\,\,,\\[2mm]
1 - r = \dfrac{\rho_{\mi}}{\epsilon} \,t + i\,
\dfrac{\nu_{\pl}}{\epsilon} \,z\,
\tilde{t}\,\,,\\[2mm]
\tilde{r} = i \,\dfrac{\rho_{\mi}}{\epsilon} \,w \,t -
\dfrac{\nu_{\pl}}{\epsilon} \,\tilde{t}\,\,,
\end{array} 
\]
which gives
\begin{equation}
\label{rtza}
\begin{array}{llll}
t = \dfrac{2 \epsilon}{\epsilon + \rho_{\mi}} \left[ 1 - zw\,
\dfrac{\epsilon + i \nu_{\pl}} {\epsilon + \nu_{\pl}}
\dfrac{\epsilon - i \rho_{\mi}}{\epsilon + \rho_{\mi}}
\right]^{-1}~, \\[3mm]
r = \dfrac{\epsilon - \rho_{\mi}}{2 \epsilon} \left[ 1 - zw
\,\dfrac{\epsilon - i \nu_{\pl}} {\epsilon + \nu_{\pl}}
\dfrac{\epsilon - i \rho_{\mi}}{\epsilon - \rho_{\mi}}
\right] t~,\\[3mm]
\tilde{t} = - \dfrac{\epsilon - i \rho_{\mi}}{\epsilon + \nu_{\pl}} \,w\, t~, \\[3mm]
\tilde{r} = \dfrac{\nu_{\pl} + i \rho_{\mi}}{\epsilon + \nu_{\pl}}
\,w\, t~.
\end{array}
\end{equation}
We have determined the  stationary states of a particle in the
presence of a quaternionic step potential for plane waves of
energy $E> V_{\0}=\sqrt{V_{\1}^{^{\2}} + V_{\2}^{^{\2}} +
V_{\3}^{\2}}$. In Figure 1, we plot, for different ratios of the
complex  and pure quaternionic potential
($V_{\1}/\sqrt{V_{\2}^{^{\2}} + V_{\3}^{^{\2}}}$\,), the four real
component of $\Phi(x)$ versus the adimensional space variable
$\sqrt{2mV_{\0}}\,x/\hbar$.  In region I, only the complex part of
$\Phi(x)$ presents an oscillatory behavior. The pure quaternionic
part decreases exponentially due to the presence of the evanescent
wave $e^{\epsilon x}$. In region II, we have a new  oscillatory
pure quaternionic wave. It is also important to note here that
increasing the value of pure quaternionic potential, we smooth the
phase changes expected in the potential region  by standard
quantum mechanics. Thus, we can conclude that this change in the
phase is caused by the complex part of the quaternionic potential.

These plane waves do not represent a physical state for a
localized incoming  particle. They have to be linearly superposed
to form wave packets. It is not our purpose to introduce the wave
packet treatment for quaternionic wave functions in this paper.
This topic deserves a deeper analysis and is, currently,  under
investigation. Nevertheless, a simple discussion  can done at this
stage. By using the stationary phase method\cite{COHEN}, we can
follow the maximum of the reflected and transmitted wave packets.
The use of a real modulation function $g(\epsilon)$ implies that
the incident wave packets reaches the point $x=0$ at $t=0$. Any
phase in the reflected and/or transmitted waves will introduce a
shift in time. To clarify this point, it can be useful to rewrite
the reflection and transmission coefficients in temrs of their
modulus and phases. By simple algebraic manipulations,  we find
\begin{eqnarray}
\label{rc}
 r & = &\frac{ (\epsilon -
\rho_{\mi})(\epsilon + \nu_{\pl}) - zw \,(\epsilon^{\2} -
\rho_{\mi}\nu_{\pl})+ i\,  zw \,  \epsilon \,  (
\rho_{\mi}+\nu_{\pl})}{ (\epsilon + \rho_{\mi})(\epsilon +
\nu_{\pl}) - zw \,(\epsilon^{\2} + \rho_{\mi}\nu_{\pl}) + i\, zw
\, \epsilon \,  ( \rho_{\mi}-\nu_{\pl})}  \nonumber \\
& = &\sqrt{\frac{ [(\epsilon - \rho_{\mi})(\epsilon + \nu_{\pl}) -
zw \,(\epsilon^{\2} - \rho_{\mi}\nu_{\pl})]^{^{\2}} + z^{\2}w^{\2}
\epsilon^{\2} ( \rho_{\mi}+\nu_{\pl})^{^{\2}} }{ [(\epsilon +
\rho_{\mi})(\epsilon + \nu_{\pl}) - zw \,(\epsilon^{\2} +
\rho_{\mi}\nu_{\pl})]^{^{\2}} + z^{\2}w^{\2}\epsilon^{\2} (
\rho_{\mi}-\nu_{\pl})^{\2} }}\, \,\, \exp[i\,(\theta_n -
\theta_d)]
\end{eqnarray}
and
\begin{eqnarray}
\label{tc}
 t 
 & = & \frac{2\epsilon}{\epsilon + \rho_{\mi}}\,\left[ \frac{
(\epsilon+\rho_{\mi})(\epsilon+\nu_{\pl})}{(\epsilon+\rho_{\mi})
(\epsilon+\nu_{\pl})- zw \,(\epsilon - i \rho_{\mi}) (\epsilon +
i \nu_{\pl})} \right] \nonumber \\
  & = &  \frac{2\epsilon (\epsilon+\nu_{\pl})}{\sqrt{[(\epsilon+\rho_{\mi})
  (\epsilon+\nu_{\pl})-zw \,(\epsilon^{\2} -
\rho_{\mi}\nu_{\pl})]^{^{\2}} + z^{\2}w^{\2} \epsilon^{\2} (
\rho_{\mi}+\nu_{\pl})^{^{\2}}}} \,\, \exp[-i\,\theta_d]
\end{eqnarray}
where
\begin{eqnarray}
\theta_n & = &  \arctan \left[ \frac{ zw\,\epsilon\, (
\rho_{\mi}+\nu_{\pl}) }{(\epsilon - \rho_{\mi})(\epsilon +
\nu_{\pl}) - zw \,(\epsilon^{\2} - \rho_{\mi}\nu_{\pl})}\right]~,
\\
\theta_d & = &\arctan \left[ \frac{ zw\,\epsilon\, (
\rho_{\mi}-\nu_{\pl}) }{(\epsilon + \rho_{\mi})(\epsilon +
\nu_{\pl}) - zw \,(\epsilon^{\2} + \rho_{\mi}\nu_{\pl})}\right]~.
\end{eqnarray}
The phases of the reflected and transmitted waves  are then given
by
\begin{eqnarray}
\theta_{r}(\epsilon;x,t) & = & \theta_n(\epsilon) -
\theta_d(\epsilon) - \epsilon \,x -
\frac{\,\,\hbar\,\epsilon^{\2}}{2\,m}\, t~,\nonumber \\
\theta_{t}(\epsilon;x,t) & = & - \theta_d(\epsilon) +
\rho_{\mi}(\epsilon) \,x - \frac{\,\,\hbar\,\epsilon^{\2}}{2\,m}\,
t~.
\end{eqnarray}
The stationary phase method suggests that the maximum of the
reflected and transmitted waves is found at the point $x=0$ for
the following time values:
\begin{eqnarray}
\tau_r & = & \frac{m}{\hbar}\,\frac{\theta'_n(\epsilon_{\0}) -
\theta'_d(\epsilon_{\0})}{\epsilon_{\0}}\,\,, \nonumber \\
\tau_t & = &
-\,\frac{m}{\hbar}\,\frac{\theta'_d(\epsilon_{\0})}{\epsilon_{\0}}\,\,,
\end{eqnarray}
where $\epsilon_{\0}$ is the maximum of the modulation function
$g(\epsilon)$. In this energy zone ($E>V_{\0}$), an immediate {\em
qualitative} difference between complex and quaternionic quantum
mechanics is found. For quaternionic potential the reflection and
transmission are {\em not} instantaneous. We shall come back to
this point later.

\subsection*{$\bullet$  REGION II - ZONE A:  COMPLEX LIMIT}

Performing the complex limit, $V_{\2,\3}\to 0$, we obtain
\begin{eqnarray*}
\nu_{\pl} & \to & \sqrt{2m\,(E + V_{\1})}\,/\hbar\,\,,\\
\rho_{\mi} & \to & \sqrt{2m\,(E - V_{\1})}\,/\hbar\,\,,\\
z,w & \to &  0\,\,.
\end{eqnarray*}
From Eq.(\ref{rtza}), we find the reflection and transmission
coefficient of standard (complex) quantum mechanics
\begin{eqnarray*}
t_{\Co} & = & 2\,\sqrt{E}\,\mbox{\Large $/$}\,\left(\sqrt{E}+\sqrt{E-V_{\1}}\right)\,\,,\\[2mm]
r_{\Co} & = &\left(\sqrt{E}-\sqrt{E-V_{\1}}\right)\,\mbox{\Large
$/$}\,\left(\sqrt{E}+\sqrt{E-V_{\1}}\right)\,\,,\\[2mm]
\tilde{r}_{\Co},\tilde{t}_{\Co} & = & 0~.
\end{eqnarray*}
Due to the real nature of $r_{\Co}$ and $t_{\Co}$, we find
instantaneous reflection and transmission. This means that at time
zero, the maximum of the incident, reflected and transmitted waves
are at $x=0$.

\subsection*{$\bullet$ REGION II - ZONE B: CONTINUITY}

Matching the continuity conditions at $x = 0$, we obtain
\begin{eqnarray}
t & = & \frac{2\epsilon}{\epsilon + i\, \nu_{\mi}}\,\left[\,1-\,
z\,w\, \frac{\epsilon +\nu_{\mi}}{\epsilon+i\,\nu_{\mi}}\,
\frac{\epsilon + i\nu_{\pl}}{\epsilon+\nu_{\pl}}\,\right]^{\mi \, \1}~,
\nonumber \\
r & = & \frac{\epsilon - i\,\nu_{\mi}}{2\epsilon}\,\left[\,1-\,
z\,w\, \frac{\epsilon +\nu_{\mi}}{\epsilon-i\,\nu_{\mi}}\,
\frac{\epsilon - i\nu_{\pl}}{\epsilon+\nu_{\pl}}\,\right]\, t~,\\
\tilde{t} & = &-\, \frac{\epsilon +
\nu_{\mi}}{\epsilon+\nu_{\pl}}\,\,\,w\,t~,  \nonumber \\
\tilde{r} & = &\frac{\nu_{\pl} -
\nu_{\mi}}{\epsilon+\nu_{\pl}}\,\,\,w\,t~. \nonumber
\end{eqnarray}
In Figure 2, we plot the four real component of $\Phi(x)$ versus
the adimensional space variable $\sqrt{2mV_{\0}}\,x/\hbar$. Zone B
is characterized by
$\sqrt{V_{\2}^{^{\2}}+V_{\3}^{^{\2}}}<E<V_{\0}$. In Figure 2, we
have considered the case $E=V_{\0}/\sqrt{2}$. Consequently, the
behavior of the stationary waves in this zone is given by the
plots corresponding to
$\sqrt{V_{\2}^{^{\2}}+V_{\3}^{^{\2}}}/\,V_{\0}<1/\sqrt{2}$. In
this zone, due to the presence of evanescent exponentials in the
transmitted waves, we find a non-zero probability to find the
particle in the region of space where $x$ is positive only for
short times. The stationary phase method can be applied to the
reflected wave.  The coefficient $r$ can be rewritten as follows
\begin{eqnarray}
r & = & \frac{ \epsilon\,[(\epsilon + \nu_{\pl})-zw\,(\epsilon +
\nu_{\mi})] + i \,[ zw \,\nu_{\pl}  ( \epsilon+\nu_{\mi}) -
\nu_{\mi} (\epsilon + \nu_{\pl})] }{\epsilon\,[(\epsilon +
\nu_{\pl})-zw\,(\epsilon + \nu_{\mi})] - i \,[ zw \,\nu_{\pl}  (
\epsilon+\nu_{\mi}) - \nu_{\mi} (\epsilon + \nu_{\pl})]
}=\exp[2\,i\,\theta]\,\,,
\end{eqnarray}
where
\begin{eqnarray*}
\theta  & = &  \arctan \left[ \frac{zw \,\nu_{\pl}  ( \epsilon +
\nu_{\mi}) - \nu_{\mi} (\epsilon + \nu_{\pl})}{
\epsilon\,(\epsilon + \nu_{\pl}) - z w \,\epsilon\,(\epsilon +
\nu_{\mi})} \right]\,\,.
\end{eqnarray*}
The phase of the reflected wave is then given by
\begin{eqnarray}
\theta_{r}(\epsilon;x,t) & = & 2\,\theta(\epsilon) - \epsilon \,x
- \frac{\,\,\hbar\,\epsilon^{\2}}{2\,m}\,t~.
\end{eqnarray}
It is important to observe that here we have a {\em quantitative}
difference between complex and quaternionic quantum mechanics.
Indeed, also for the standard quantum theory the reflection is not
instantaneous. The reflect wave reaches $x=0$ at time
\begin{eqnarray}
\tau_r & = &2\,
\frac{m}{\hbar}\,\frac{\theta'(\epsilon_{\0})}{\epsilon_{\0}}\,\,.
\end{eqnarray}

\subsection*{$\bullet$ REGION II - ZONE B: COMPLEX LIMIT}

Performing the complex limit, $V_{\2,\3}\to 0$, we obtain

\begin{eqnarray*}
t & \to & t_{\Co}=2\,\sqrt{E}\,\mbox{\Large
$/$}\,\left(\sqrt{E}+i\,\sqrt{V_{\1}-E}\right)\,\,,\\
r & \to & r_{\Co}=
\left(\sqrt{E}-i\,\sqrt{V_{\1}-E}\right)\,\mbox{\Large
$/$}\,\left(\sqrt{E}+i\,\sqrt{V_{\1}-E}\right)\,\,,\\
\tilde{r},\tilde{t}  & \to &  0~.
\end{eqnarray*}
The presence of the phase
\[
\theta_{\Co}  =  \arctan \left( - \sqrt{ \dfrac{V_{\1} - E}{E} }
\right)
\]
in $r_{\Co}$ shows that the reflection is not instantaneous. The
delay time\cite{COHEN} is
\begin{eqnarray}
\tau_{r,{\Co}} & = &2\,
\frac{m}{\hbar}\,\frac{\theta'_{\Co}(\epsilon_{\0})}{\epsilon_{\0}}=
\frac{\hbar}{E_{\0}}\,\sqrt{ \dfrac{E_{\0}}{V_{\1} - E_{\0}}}\,\,.
\end{eqnarray}

\subsection*{$\bullet$ REGION II - ZONE C: CONTINUITY}

The last zone is characterized by  $E < \sqrt{V_{\2}^{^{\2}} +
V_{\3}^{^{\2}}}$. Continuity conditions give
\begin{eqnarray}
t & = & \frac{2\epsilon}{\epsilon + \sigma_{\mi} + i\,
\sigma_{\pl} }\,\left[\,1-\, z\,w\, \frac{\epsilon  + \sigma_{\pl}
- i \,\sigma_{\mi}}{\epsilon+ \sigma_{\mi} + i\, \sigma_{\pl}}\,\,
\frac{\epsilon - \sigma_{\mi} + i \,\sigma_{\pl}}{\epsilon+
\sigma_{\pl} + i\,
\sigma_{\mi}}\,\right]^{\mi \, \1}\,\,,  \nonumber\\
r & = & \frac{\epsilon - \sigma_{\mi} - i\,
\sigma_{\pl}}{2\epsilon}\,\left[\,1-\, z\,w\, \frac{\epsilon +
\sigma_{\pl} - i \,\sigma_{\mi}}{\epsilon-\sigma_{\mi} - i\,
\sigma_{\pl}}\,\, \frac{\epsilon +  \sigma_{\mi} - i\,
\sigma_{\pl} }{\epsilon+ \sigma_{\pl} + i\,
\sigma_{\mi} }\,\right]\, t\,\,,\\
\tilde{t} & = &-\, \frac{\epsilon + \sigma_{\pl} - i\,
\sigma_{\mi}}{\epsilon+ \sigma_{\pl} + i\,
\sigma_{\mi}}\,\,\,w\,t\,\,,  \nonumber \\
\tilde{r} & = &\frac{ 2 \,i \,\sigma_{\mi}}{\epsilon+ \sigma_{\pl}
+ i \,\sigma_{\mi} }\,\,\,w\,t \,\,.  \nonumber
\end{eqnarray}
The stationary waves corresponding to this case are also shown in
Figure 2, see the plots with
$\sqrt{V_{\2}^{^{\2}}+V_{\3}^{^{\2}}}/\,V_{\0}>1/\sqrt{2}$. To
discuss the time reflection, we rewrite $r$ in terms of its
modulus and phase,
\begin{eqnarray}
 r & = & \frac{(\epsilon - \sigma_{\mi} - i\,
\sigma_{\pl})(\epsilon + \sigma_{\pl} + i \,\sigma_{\mi}) \,e^{i
\varphi} - (\epsilon + \sigma_{\pl} - i\, \sigma_{\mi}) (\epsilon
+ \sigma_{\mi} - i\, \sigma_{\pl})\, e^{- i \varphi}}{(\epsilon +
\sigma_{\pl} + i \,\sigma_{\mi})(\epsilon + \sigma_{\mi} + i\,
\sigma_{\pl}) \,e^{i \varphi} - (\epsilon + \sigma_{\pl} - i\,
\sigma_{\mi})(\epsilon - \sigma_{\mi} + i \,\sigma_{\pl}) \,e^{- i
\varphi}} = \nonumber\\
 & = & \exp[\,2 \,i\,\theta]
\end{eqnarray}
where
\begin{eqnarray*}
 \theta & = & \arctan \left[\, \frac{\epsilon \,(\epsilon +
 \sigma_{\pl})\, \tan \varphi + \epsilon \,\sigma_{\mi}}{\epsilon
 \,\sigma_{\mi}-
 (\epsilon \,\sigma_{\pl} + \sigma_{\mi}^2 +
 \sigma_{\pl}^2 )\, \tan \varphi}\,\right]~.
\end{eqnarray*}
In this zone, we do not have a complex limit case. It is important
to observe that a {\em new} phenomenon appears. The oscillatory
behavior of the particle in region II is damped due to the
presence  of the evanescent wave $e^{- \sigma_{\pl} x}$. Thus, a
non-zero probability to find the particle in the potential region
only exists for short times.

\section*{V. RELATIONSHIPS BETWEEN COMPLEX AND QUATERNIONIC QUANTUM MECHANICS.}

In the last years, the Schr\"odinger equation in the  presence of
quaternionic (constant) potentials has been a matter of study and
discussion in literature. This is justified in view of a possible
understanding of the role that a quaternionic quantum theory could
play in the real physical world. As remarked by Adler\cite{ADL}
all known physical phenomena appear to be very well described by
complex quantum mechanics. Nevertheless, to see if quaternionic
quantum mechanics represents a possible way to describe the nature
or if it is only an interesting mathematical exercise, we have to
use and test this formalism in simple quantum mechanical systems.
With respect to previous works regarding   potential barrier
diffusion\cite{DAV89,DAV92,DEDUC2} and potential well bound
states\cite{DEDUC3}, in this paper, we have preferred  to go back
in our analysis of non relativistic quaternionic quantum mechanics
by studying the potential step.

The study presented in this paper can be seen as an attempt to
understand, by starting from an {\em analytic} solution of a
 simple quantum mechanical system, where and if differences between
 standard quantum mechanics and  theoretical solutions obtained by solving the
 Schr\"odinger equation in  the presence of a quaternionic step
 perturbation can be observed. The main difficulty in obtaining
 quaternionic solutions of a physical problem is due to the fact that,
 in general, the standard mathematical methods of resolution break down.
Nevertheless, the recent progress in quaternionic differential
theory\cite{DEDUC1,DEDUC1a,DEDUC1b} and linear
algebra\cite{DESCO1,DESCO2} give the possibility to use "new"
quaternionic mathematical tools. As a direct consequence  of this,
we have been able to find an analytic solution for the stationary
states in the presence of a quaternionic potential step. This
means that we have now (for the first time) the possibility of
describing in detail {\em qualitative} differences between complex
and quaternionic quantum mechanics.

\subsection*{$\bullet$ EXPERIMENTAL PROPOSALS IN QUATERNIONIC
QUANTUM MECHANICS.}

The earliest experimental proposals to test quaternionic
deviations from complex quantum mechanics\cite{PER79} suggested
that the  non commutativity of  quaternionic phases could be
observed in Bragg scattering by crystal made of three different
atoms, in neutron interferometry and in meson regeneration. In
1984, the neutron interferometric experiment was realized by
Kaiser, George and Werner\cite{KAI84}. The neutron wave function
traversing slabs of two dissimilar materials (titanium and
aluminum) should experience  the non commutativity of the phase
shifts when the order in which the barriers are traversed is
reversed. The experimental result showed that the phase shifts
commute to better than one part in $3 \times 10^{\4}$. To explain
this null result, Klein postulated\cite{KLE88} that quaternionic
potentials act only for some of the fundamental forces and
proposed an experiment for testing possible violations of the
Schr\"odinger equation by permuting the order in which nuclear,
magnetic and gravitational potentials act on neutrons in an
interferometer.

The first theoretical analysis of two quaternionic potential
barriers was developed by Davies and McKellar\cite{DAV92}. In
their paper, by translating the   quaternionic Schr\"odinger
equation into a pair of coupled complex equations and solving the
corresponding complex system by numerical methods, Davies and
McKellar showed that, notwithstanding the presence of complex {\em
instead} of quaternionic phases, the predictions of quaternionic
quantum mechanics differ from those of the usual theory. In
particular, they pointed out that differently from the  complex
quantum mechanics prediction, where the left and right
transmission amplitudes, $t_{\L}$ and $t_{\R}$, are equal in
magnitude and in phase,  in the quaternionic quantum mechanics
only the magnitudes $|t_{\L}|$ and $|t_{\R}|$ are equal. So, the
measurement of a phase shift should be an indicator of
quaternionic effects {\em and} of space dependent phase
potentials. However, this conclusion leads to the embarrassing
question of why there was no phase change in the experiment
proposed by Peres\cite{PER79} and realized by Kaiser, George and
Werner\cite{KAI84}. To reconcile the theoretical predictions with
the experimental observations, Davies and McKellar reiterated the
Klein conclusion and suggested to subject the neutron beam to
different interactions in permuted order.

In the final chapter of the Adler book~\cite{ADL}, we find and
intriguing question. Do the Kayser and colleagues experiment, and
the elaborations on it proposed by Klein actually test for
residual quaternionic effects? According to the non relativistic
quaternionic scattering theory developed by Adler~\cite{ADL} the
answer is clearly {\em no}. Experiments to detect a phase shift
are equivalent to detect {\em time reversal violation}, which so
far has not been detectable in neutron-optical experiments.

\subsection*{$\bullet$ QUATERNIONIC POTENTIAL STEP, CP VIOLATION AND KAONS SYSTEM.}

Based on the previous considerations,  experimental proposals to
test quaternionic deviations from standard quantum mechanics
should involve CP violation dynamical systems. A natural candidate
to a such investigation could be the system of
K-mesons\cite{ZUB,SAK}. The quaternionic time reversal violation
potential (see Section II)
\[ W(\r) = |W(\r)|\, \exp[\,i\,\theta(\r)\,] \]
should be directly responsible for CP violation effects. The
experimental results on $K_{\Sh,\L}$ (a $K_{\L}$ meson decays more
often to $\pi^{\mi}e^{\pl}\bar{\nu}_e$ than to
$\pi^{\pl}e^{\mi}\nu_e$\cite{DP04}) could be useful to estimate
the modulus and phase of the pure quaternionic part of this
"effective"  potential.

Once determined the magnitude of the quaternionic perturbation, by
using the analytic solution obtained in this paper, could be
possible (through a stationary phase analysis) to explicitly
calculate the reflection and transmission times of a K-meson
particle scattered by a complex potential step in the presence of
a quaternionic (CP violating) perturbation. In the case of a
above-potential incident particle the diffusion from a pure
complex potential (standard quantum mechanics) happens
instantaneously, i.e.
\[\tau_{r,{\Co}}=\tau_{t,{\Co}}=0\,\,.\]
The possibility to {\em analytically} solve the corresponding
quaternionic problem gives us the chance to see an immediate {\em
qualitative} difference between complex and quaternionic quantum
mechanics. The presence of a quaternionic potential surprisingly
modifies the reflection and transmission times which are now non
zero,
\[
\tau_r  =  \frac{m}{\hbar}\,\frac{\theta'_n(\epsilon_{\0}) -
\theta'_d(\epsilon_{\0})}{\epsilon_{\0}}\,\,\,\,\,\mbox{and}
\,\,\,\,\, \tau_t =
-\,\frac{m}{\hbar}\,\frac{\theta'_d(\epsilon_{\0})}{\epsilon_{\0}}\,\,,
\]
where $\epsilon_{\0}$ is the maximum of the wave packet modulation
function.

With this paper, we would have liked to close the debate on the
role that quaternionic potentials could play in quantum mechanics,
but more realistically, we simply contribute to the general
discussion. Physical interpretations of quaternionic solutions
still represent a {\em delicate} question and  before  proposing a
detailed experimental test, we think that  more mathematical
questions should be addressed and deeply investigated.

\subsection*{$\bullet$ COMPLEX AND QUATERNIONIC GEOMETRIES.}

To give a satisfactory probability interpretation, amplitudes of
probability must be defined in {\em associative division algebras}
\cite{ADL}. Amplitudes of probabilities defined in non-division
algebras fail to satisfy the requirement that in the absence of
quantum interference effects, probability amplitude superposition
should reduce to probability superposition. The associative law of
multiplication (which fails for the octonions) is needed to
satisfy the completeness formula and to guarantee that the
Schr\"odinger anti-self-adjoint operator leaves invariant the
inner product.

At first glance it appears that we cannot formulate quantum
theories by using wave functions defined in non-division or
non-associative  algebras. This is an erroneous conclusion because
the constraint concerns the inner product and {\em not} the kind
of Hilbert space in which we define our wave functions. Amplitudes
of probability have to be given in $\mathbb{C}$ or $\mathbb{H}$
(complex or quaternionic geometry) but vectors in the Hilbert
space have no limitation. We can formulate a consistent
complexified quaternionic \cite{CQ1,CQ2,CQ3,CQ4,CQ5} or octonionic
\cite{OCT1,OCT2} quantum mechanics by adopting complex inner
products. The use of complex inner product represents a
fundamental tool in applying a Clifford algebraic formalism to
physics and plays a fundamental role in looking for geometric
interpretation of the algebraic structures in relativistic
equations  and gauge theories \cite{ST1,ST2,ST3,ST4,ST5}. The
choice of quaternionic inner product seems to be best adapted to
investigate deviations from the standard complex theory  in
quantum mechanics \cite{SOU1,SOU2} and field theory
\cite{ADLq1,ADLq2}

\subsection*{$\bullet$ CONCLUSIONS AND OUTLOOKS.}

We conclude this paper by listing the most interesting features of
our analysis and future investigations suggested by our results.\\
(1) An {\em analytic} solution for a simple quantum mechanical
system (quaternionic potential step) has been given
(previous studies on the quaternionic Schr\"odinger equation
have been performed by numerical calculations).\\
(2) Our plane wave analysis immediately show {\em qualitative}
differences between complex and quaternionic quantum mechanics
(see for example the reflection and transmission times for
above-potential diffusion and the oscillatory behavior in the new
region below the potential).\\
(3) A plane wave analysis for a quaternionic barrier can be now
developed by using an analytic {\em two step approach}.\\
(4) The plane wave results (valid in the physical situation of
complete interference) should be revised by introducing a
quaternionic {\em wave packet} formalism (particle viewpoint).
This should confirm and explain the reflection and transmission
times
obtained by the stationary phase method.\\
(5) In the quaternionic barrier analysis, we expect {\em
qualitative} differences between complex and quaternionic quantum
dynamical system.

(5.1) For  above-potential diffusion, the quaternionic wave
packets will be characterized by {\em new} reflection and
transmission times with respect to the standard (complex)
case\cite{MUL1,MUL2}.

(5.2) In the tunnelling zone, the quaternionic  {\em Hartman}
effect have to be investigated and confronted with the standard
one which predicts (for a long barrier) instantaneous
transmission\cite{HAR1,HAR2,HAR3}.

(5.3) In the {\em new} below-potential region, a {\em Klein-like}
phenomenon\cite{KLE1,KLE2} appears and it should be interpreted
within a
non-relativistic context.\\
(6) The natural candidate to quaternionic experimental proposals
seems to be the system of $K$-mesons. The above suggested
investigations should give a more clear idea about the possibility
to {\em really} perform an experiment involving non-relativistic
oscillating particles and CP violating potential barriers. This
probably should close the debate on the use  of a quaternionic
mathematical formalism in quantum theories.



\newpage

\begin{figure}[hbp]
\hspace*{-2.5cm}
\includegraphics[width=19cm, height=22cm, angle=0]{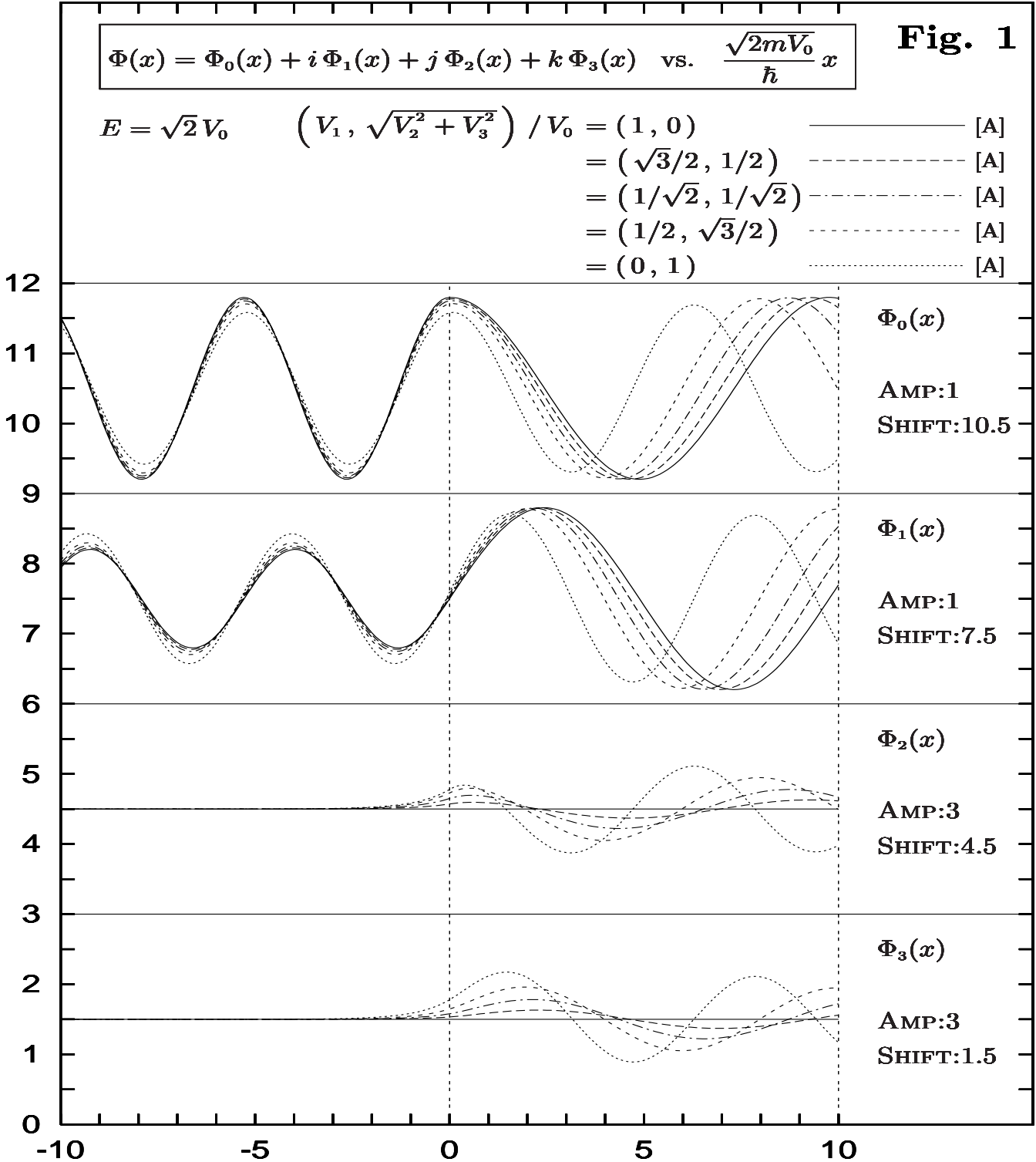}
\caption{Space is divided into region I ($x<0$) and region II
($x>0$). There is a constat quaternionic potential
($iV_{\1}+jV_{\2}+kV_{\3}$) in region II whereas in region I there
is no potential. The space dependence of the quaternionic
stationary wave function $\Phi(x)$ is plotted for the energy zone
A ($E>V_{\0}=\sqrt{V_{\1}^{^{\2}}+V_{\2}^{^{\2}}+V_{\3}^{^{\2}}}$)
and for different complex/pure quaternionic potential ratios. The
plots for the complex part of $\Phi(x)$   exhibit an oscillatory
behavior both for region I and region II. The pure quaternionic
part is practically absent in the free potential region.}
\end{figure}

\newpage

\begin{figure}[hbp]
\hspace*{-2.5cm}
\includegraphics[width=19cm, height=22cm, angle=0]{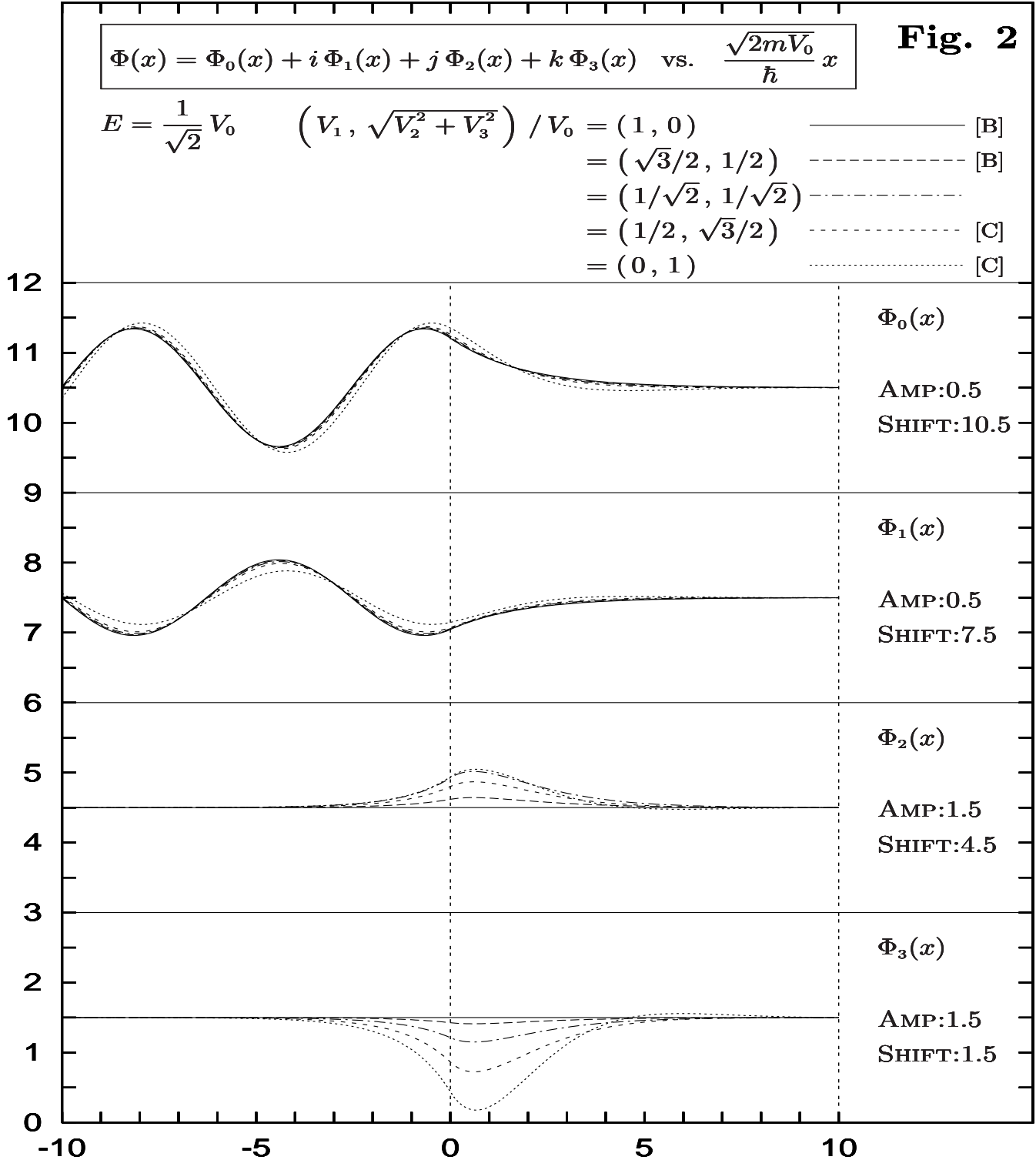}
\caption{The space dependence of the quaternionic stationary wave
function $\Phi(x)$ is plotted for the energy zone B
($\sqrt{V_{\2}^{^{\2}}+V_{\3}^{^{\2}}}<E<V_{\0}$) and C
($E<\sqrt{V_{\2}^{^{\2}}+V_{\3}^{^{\2}}}$), and for different
complex/pure quaternionic potential ratios. The plots for the
complex part of $\Phi(x)$   exhibit an evanescent  behavior in
region II. The pure quaternionic part is practically zero far from
the discontinuity point. An interesting oscillatory behavior
smoothed by the evanescent waves is also present in region II for
energy values in zone C. This remembers the Klein zone in the
Dirac equation.}
\end{figure}


\begin{thebibliography}{99}

\bibitem{ADL}
S. L. Adler, {\it Quaternionic quantum mechanics and quantum
fields}, (New York: Oxford University Press, 1995).


\bibitem{DAV89}
A. J. Davies and B. H. McKellar, ``Non-relativistic quaternionic
quantum mechanics'', Phys. Rev. A {\bf 40},  4209--4214 (1989).

\bibitem{DAV92}
A. J. Davies and B. H. McKellar, ``Observability of quaternionic
quantum mechanics'', Phys. Rev. A {\bf 46}, 3671--3675 (1992).




\bibitem{DEDUC1}
S, De Leo and G. C. Ducati , ``Quaternionic differential
operators´´, J. Math. Phys. {\bf 42}, 2236--2265 (2001).

\bibitem{DEDUC1a}
S, De Leo and G. C. Ducati , ``Solving simple quaternionic
differential equations´´, J. Math. Phys. {\bf 44}, 2224-2233
(2003).

\bibitem{DEDUC1b}
S, De Leo and G. C. Ducati , ``Real linear quaternionic
differential operators´´, Comp. Math. Appl. {\bf 48}, 1893-1903
(2004).


\bibitem{DEDUC2}
S. De Leo, G. Ducati and C. Nishi, ``Quaternionic potential in
non-relativistic quantum mechanics´´, J. Phys. A {\bf 35},
5411--5426 (2002).

\bibitem{DEDUC3}
S. De Leo and G. C. Ducati, ``Quaternionic bound states´´, J.
Phys. A {\bf 38}, 3443--3454 (2005).






\bibitem{COHEN}
C. Choen-Tannoudji, B. Diu and F. Lal\"oe, {\it Quantum
mechanics}, (New York: John Wiley $\&$ Sons, 1977).



\bibitem{MAT}
P. T. Matthews, {\it Introduction to quantum mechanics}, (New
York: McGraw-Hill, 1963).


\bibitem{MER}
E. Merzbacher, {\it Quantum mechanics}, (New York: John Wiley $\&$
Sons, 1970).


\bibitem{DESCO1}
S. De Leo and G. Scolarici, "Right eigenvalue equation in
quaternionic quantum mechanics", J. Phys. A {\bf 33}, 2971--2995
(2000).

\bibitem{DESCO2}
S. De Leo, G. Scolarici and L. Solombrino, "Quaternionic
eigenvalue problem", J. Math. Phys. {\bf 43}, 5812--2995 (2002).


\bibitem{PER79}
A. Peres, ``Proposed test for complex versus quaternion quantum
theory'', Phys. Rev. Lett. {\bf 42}, 683--686 (1979).

\bibitem{KAI84}
H. Kaiser, E. A. George and S. A. Werner, ``Neutron
interferometric search for quaternions in quantum mechanics'',
Phys. Rev. A {\bf 29}, 2276--2279 (1984).


\bibitem{KLE88}
A. G. Klein, ``Schr\"odinger inviolate: neutron optical searches
for violations of quantum mechanics'', Physica B {\bf 151}, 44--49
(1988).

\bibitem{ZUB}
C. Itzykson and J.B. Zuber, {\em Quantum Field Theory},
McGraw-Hill, Singapore (1985).


\bibitem{SAK}
J.J. Sakurai, {\em Advanced Quantum Mechanics}, Addison-Wesley,
New York (1987).

\bibitem{DP04}
S. Eidelman et al., "The Review of Particle Physics", Phys. Lett.
B {\bf 592}, 1 (2004)


\bibitem{CQ1}
A.W. Conway, "Quaternion treatment of the electron wave equation",
Proc. Roy. Soc. A {\bf 162}, 145–154 (1937).


\bibitem{CQ2}
A.W. Conway, "Quaternions and quantum mechanics", Acta Pont. Acad.
Scien. {\bf 12} 259–277 (1948).


\bibitem{CQ3}
S. De Leo, "One component Dirac equation", Int. J. Mod. Phys. A
{\bf 11}, 3973-3985 (1996).


\bibitem{CQ4}
S. De Leo and W. A. Rodrigues, "Quaternionic electron theory",
Int. J. Theor. Phys. {\bf 37}, 1511-1529 (1998); {\em ibidem}
1707-1720 (1998).

\bibitem{CQ5}
S. De Leo, W. A. Rodrigues and J. Vaz, "Complex geometry and Dirac
equation", Int. J. Theor. Phys. {\bf 37}, 2415-2431 (1998).


\bibitem{OCT1}
S. De Leo and K. Abdel-Khalek, "Octonionic quantum mechanics and
complex geometry", Prog. Theor. Phys. {\bf 96}, 823-831 (1996);
"Octonionic Dirac equation",  {\em ibidem}  833-845 (1996).

\bibitem{OCT2}
S. De Leo and K. Abdel-Khalek, "Octonionic representations of
GL(8,R) and GL(4,C)", J. Math. Phys. {\bf 38}, 582-598 (1997).



\bibitem{ST1}
D. Hestenes, {\em Spacetime algebra}  (New York: Gordon and
Breach, 1966).

\bibitem{ST2}
A. Lasenby, C. Doran, and S. Gull, "A multivector derivative
approach to Lagrangian field theory", Found. Phys. {\bf 23},
1295–1327 (1993).

\bibitem{ST3}
S. Gull, C. Doran, and A. Lasenby, "Electron paths, tunneling and
diffraction in the spacetime algebra", Found. Phys. {\bf 23},
1329–1356 (1993).





\bibitem{ST4}
S. De Leo, Z. Oziewicz, W.A. Rodrigues, and J. Vaz, "Dirac
Hestenes Lagrangian", Int. J. Th. Phys. {\bf 38}, 2349–2369
(1999).


\bibitem{ST5}
A. Lasenby and J. Lasenby, "Applications of geometric algebra in
physics and links with engineering", 430–457,  in E.B. Corrochano
and G. Sobczyk, {\em Geometric Algebra with Applications in
Science and Engineering} (Boston: Birkhauser, 2001) .







\bibitem{SOU1}
J. Soucek, Quaternion quantummechanics as a true 3+1dimensional
theory of tachyons, J. Phys. A 14 (1981) 1629–1640.


\bibitem{SOU2}
J. Soucek, Quaternion quantum mechanics as a description of
tachyons and quarks, Czech. J. Phys. B 29 (1979) 315–318.



\bibitem{ADLq1}
S.L. Adler, "Quaternionic quantum field theory", Phys. Rev. Lett.
{\bf 55},  783–786 (1985).


\bibitem{ADLq2}
S.L. Adler, "Quaternionic quantum field theory", Comm. Math. Phys.
{\bf 104}, 611–656 (1986).






\bibitem{MUL1}
A. Anderson, "Multiple scattering approach to one-dimensional
potential problem", Am.  J. Phys. {\bf 57},    230-235  (1989).

\bibitem{MUL2}
A. Bernardini, S. De Leo and P. Rotelli, "Above barrier potential
diffusion", Mod. Phys. Lett. A {\bf 19}, 2717-2725 (2004).

\bibitem{HAR1}
T.E. Hartman, "Tunnelling of a wave packet", J. Appl. Phys. {\bf
33}, 3427-3432 (1962).

\bibitem{HAR2}
A. Pablo, L. Barbero, H.E. Hern\'andez-Figueroa, and E. Recami,
"Propagation speed of evanescent modes", Phys. E {\bf 62},
8628-8635 (2000).

\bibitem{HAR3}
V.S. Olkhovsky, E. Recami, and J, Jakiel, "Unified time analysis
of photon and particle tunnelling",  Phys. Rep. {\bf 398}, 133-178
(2004).



\bibitem{KLE1}
O. Klein, "Die Reflexion von Elektronrn an einem Potentialsprun
nach der  relativistichen Dynamik von Dirac", Z. Phys. {\bf 53},
157-165 (1929).

\bibitem{KLE2}
M. Soffel, B. M\"uller, and W. Greiner, "Stability and decay of
the Dirac vacuum in external gauge fields", Phys. Rep. {\bf 85},
51-122 (1982).



\end{thebibliography}
\end{document}